\documentclass[superscriptaddress,12pt,notitlepage]{revtex4-1} 
\usepackage{amsmath,amssymb,amsfonts,graphicx,bbm}
\usepackage{color}
\setcitestyle{super}
\usepackage{amscd}
\usepackage{dsfont}
\usepackage{euscript}
\usepackage{graphicx}
\usepackage{color}

\newcommand{\abs}[1]{\left\vert#1\right\vert}

\renewcommand{\i}{\textrm{i}}

\begin{document}
\title[Relaxation processes in action signals]{The nature of relaxation processes revealed by the action signals of phase modulated light fields.}

\author{Vladimir Al. Osipov}\email{Vladimir.Al.Osipov@gmail.com}
\affiliation{Chemical Physics, Lund University, Getingev\"agen 60, 222 41, Lund, Sweden}
\author{Xiuyin Shang}\affiliation{Chemical Physics, Lund University, Getingev\"agen 60, 222 41, Lund, Sweden}
\affiliation{Agricultural University of Hebei, Lingyusi 289, 071001 Baoding, Hebei, China}
\author{Thorsten Hansen}\affiliation{Chemistry Department, University of Copenhagen, Universitetspaken 5, DK-2100, Copenhagen, Denmark}
\author{T\~onu Pullerits}\affiliation{Chemical Physics, Lund University, Getingev\"agen 60, 222 41, Lund, Sweden}
\author{Khadga Jung Karki}\affiliation{Chemical Physics, Lund University, Getingev\"agen 60, 222 41, Lund, Sweden}
\date {\today}

\begin{abstract}  We introduce a generalized theoretical approach to study action signals induced by the  absorption of two-photons from two phase modulated laser beams and subject it to experimental testing for two types of photoactive samples, solution of rhodamine 6G and GaP photodiode. In our experiment, the phases of the laser beams are modulated at the frequencies $\phi_1$ and $\phi_2$, respectively. The action signals, such as photoluminescence and photocurrent, which result from the absorption of two photons, are isolated at frequencies~$m\phi$ ($\phi=\abs{\phi_1-\phi_2}$, $m=0,1,2,\dots$). We demonstrate that the ratio of the amplitudes of the secondary ($m=2$) and the primary ($m=1$) signals, $A_{2\phi}:A_{\phi}$, is sensitive to the type of relaxation process taken place in the system and thus can be used for its identification. Such sensitivity originates from cumulative effects of non-equilibrated state of the system between the light pulses. When the cumulative effects are small, i.e. the relaxation time is much shorter then the laser repetition rate or the laser intensity is high enough to dominate the system behavior, the ratio achieves its reference value $1:4$ (the signature of two-photon absorption). In the intermediate regimes the ratio changes rapidly with the growth of intensity from zero value in case of second order relaxation process, while it demonstrates slow monotonic decrease for linear relaxation. In the article we also determine the value of the recombination rate in a GaP photodiode by using the above approach.  

\end{abstract}
\maketitle
\setlength{\textfloatsep}{20pt plus 4pt minus 4pt}
\setlength{\floatsep}{20pt plus 4pt minus 4pt}
\setlength{\columnsep}{9pt}

\section{Introduction}
Intensity-modulation of continuous laser beams have been commonly used in the measurement of life-times of various action signals, such as photoluminescence (PL)~\cite{JGH1984,CLEGG_1993,LAKOWICZ_2006} and photocurrent.~\cite{KRAUSE_2010,WILLIAMS_2011,LVOVICH_2015} The intensity modulation technique can be also implemented with pulsed lasers, wherein the peak intensity of a train of pulses are modulated. One of the advantages of pulsed excitation is that they can induce nonlinear interactions in the media due to the high peak intensity achievable within one pulse. Such modified techniques have been successfully used in multiphoton life-time imaging.~\cite{XU_2013} Among a wide variety of methods (electro-optic modulation or reflection from vibrating surfaces) that can be used to modulate the intensity of the laser beams, only a few, such as the interference of two phase modulated beams in a Mach-Zehnder interferometer,~\cite{MARCUS_2006,KARKI_2014C,KARKI_2016A} have been shown to generate a clean modulation (without undesirable sidebands at the multiples of the modulation frequency) of intensity at a single frequency. Single frequency modulated light-fields have recently been used in phase synchronous detection of different coherent and incoherent nonlinear signals.~\cite{KARKI_2016A,STEINKEMEIER_2015} In our experiment,~\cite{KARKI_2016A} two beams, whose phases are modulated at $\phi_1$ and $\phi_2$, are used to excite PL from a fluorophore. Collinear combination of the two phase modulated beams leads to the modulation of the total intensity at the frequency $\phi=\abs{\phi_2-\phi_1}$. 

In a typical light-matter interaction, the absorption of intensity modulated light modulates the perturbations of the sample. Consequently, the response from the sample, also known as the action signal, is modulated. However, as the response is stretched in time, one typically observes a phase lag and demodulation in the action signal relative to the perturbation.~\cite{JGH1984,CLEGG_1993,LAKOWICZ_2006} Conventional techniques use these information to measure the life-time when the signal decay can be described by a single-exponential. If the signal has multi-exponential decay components, one typically measures the response at multiple modulation frequencies in order to estimate the life-times of each decay components. In all cases the measurements are carried out at a constant average excitation intensity. On the other hand, the relaxation processes can be non-exponential and their relative contributions can depend on the strength of the perturbation. Investigation of such relaxation processes by the measurement of life-times (in both frequency-domain as well as time domain measurements) remains a challenge. Here, we show that the demodulation of the action signal as a function of the excitation intensity (or the strength of perturbation) can be used to distinguish the different relaxation processes. We use the ratio of the action signals at two different frequencies, $\phi$ and $2\phi$, as the observable and provide detailed theoretical analysis of the ratio dependence on the experimental parameters, such as excitation intensity, and system parameters, such as life-time of the response. In particular, we analyze the response from molecules and semiconductors that are perturbed by two-photon absorption of a train of laser pulses whose peak intensity is modulated at a single frequency, $\phi$. The two-photon absorption process perturbs the system at two frequencies, $2\phi$ and $\phi$, with a well defined ratio of 1:4.~\cite{KARKI_2016A} We show that the ratio of the PL signals at $2\phi$ and $\phi$, $A_{2\phi}:A_{\phi}$, from molecules, which are excited by two-photon absorption, is also close to 1:4. In the case of molecules that have long lived PL with mono-exponential decay, the ratio decreases with the increase in the excitation intensity. On the other hand, if the PL lifetime is very short compared to the time interval between the laser pulses, the ratio does now show significant change. Our experimental results of two-photon PL from Rhodamine 6G support the theory.  Moreover, our results show that the change in the ratio as a function of the excitation intensity shows characteristic features that can be used to distinguish different relaxation processes. We have analyzed the ratio of photo-current signals from a semiconductor in which the relaxation is a second order process. We observe that the ratio shows characteristic shape with minima at low excitation intensity. The position of the minima depends on  the recombination rate, and we use this information to  quantify the recombination of the charge carriers in a GaP photodiode that is excited by two-photon absorption of femtosecond pulses at 800 nm.

\section{Materials and methods}
\subsection{Theory} At the first step of our theoretical analysis of phase modulated multiphoton action signals, we assume that the sample is an ensemble of two-level systems that are placed in a homogeneous surrounding. Let $P(t)$ be the population in the first excited state $S_1$ at time $t$. If the depopulation of $S_1$ is by a simple stochastic relaxation to the ground state $S_0$ (Fig.1 a), then the evolution of $P(t)$ follows the equation 
\begin{equation}\label{continiousdecay}
\frac{dP(t)}{dt}=-\Gamma P(t)+ R(t),
\end{equation}
where $\Gamma$ is the rate of relaxation (depopulation) and $R(t)$ is the rate of  $S_0 \rightarrow S_1$ transition due to interaction with the light of intensity $I(t)$. The model described by Eq.\eqref{continiousdecay} assumes that the intensity of radiation is small (i.e. $R(t)$ is small), such that $P(t)\ll 1$. 

Let the sample be excited by light which intensity is modulated at frequency $\phi$. In case of linear absorption the transition rate is then oscillates with the same frequency,
\begin{equation}\label{intensitymodulation}
R(t) = s I(t)=R\big(1+A_I\sin \phi t\big),\qquad R \equiv s I_0,
\end{equation}
where $s$ is the absorption cross-section, and $I$ is measured in number of photons coming in one second per unit square, i.e. the units are photon/(s cm$^2$). The sinusoidal modulation of the transition rate leads to the modulation of the excited state population: 
 \begin{equation}\label{fluorescencemodulation}
 P(t)\propto I_0\tau\big(1+A_F\sin(\phi t - \Phi)\big),
 \end{equation}
 where $A_F$ and $\Phi$ are the amplitude and the phase shift in the modulation of the excited state population, respectively. The modulation of the excited state population in turn leads to the modulation of the PL; $F(t)\propto P(t)$. Substitution of Eq.\eqref{intensitymodulation} and \eqref{fluorescencemodulation} in Eq.\eqref{continiousdecay} gives~\cite{JGH1984} 
 \begin{equation}\label{amplitudescontinious}
 A_F=A_I/\sqrt{1+\phi^2\tau^2},\qquad \sin \Phi=\phi\tau/\sqrt{1+\phi^2\tau^2},\qquad\cos\Phi=1/\sqrt{1+\phi^2\tau^2}.
 \end{equation}
 In practice, analysis of the signal $F(t)$ can be done with the help of Fourier transform,
 \begin{multline}\label{Fourier}
 \mathcal{F}[P(t)](\omega)=\frac{1}{2\pi}\int_{-\infty}^\infty P(t)\;e^{-\i\omega t}dt=I_0\tau\delta(\omega)\\+\frac{I_0\tau A_F}{2}\delta(\omega-\phi)(\sin\Phi+\i\cos\Phi)+\frac{I_0\tau A_F}{2}\delta(\omega+\phi)(\sin\Phi-\i\cos\Phi).
 \end{multline}
 The value of $\tan\Phi$ can be calculated as the ratio of real and imaginary parts of the amplitude at the peak $\omega=\phi$. The mean square of these amplitudes gives $2 A_F$. The Fourier transform of $P(t)$ and $I(t)$ are connected by the formula
 $$
 \mathcal{F}[P(t)](\omega)=\frac{\tau}{1-\i \omega\tau}\mathcal{F}[I(t)](\omega).
 $$
 Thus, the  measurements of the relative phase shift and damping allows us to calculate the PL life-time. This approach is widely used in the measurements of the PL life-time by intensity modulation of a continuous laser beam.~\cite{JGH1984,CLEGG_1993,LAKOWICZ_2006}
 
\begin{figure}
\hfill \textbf {a.}\hspace{5pt} \includegraphics[scale=0.15]{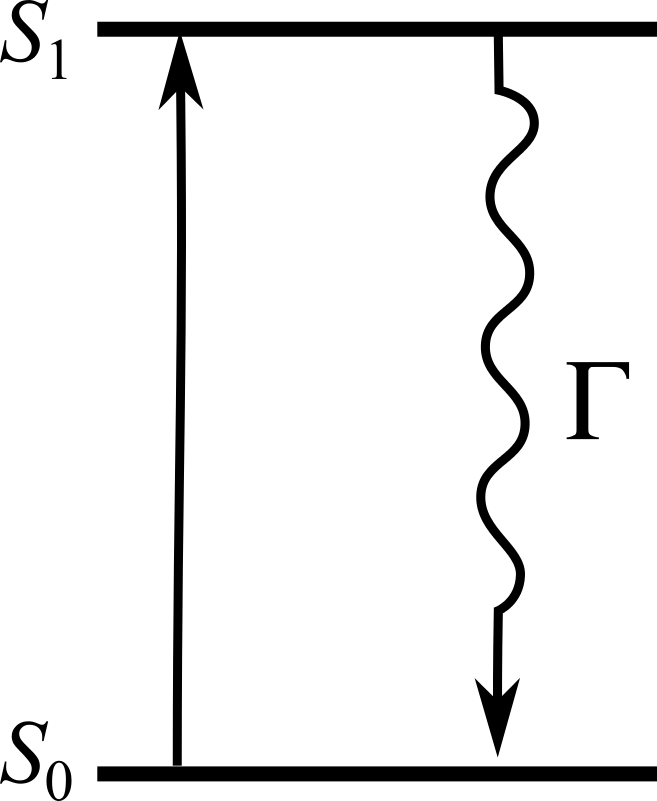}\hfill
\textbf{b.}\hspace{5pt} \includegraphics[scale=0.45]{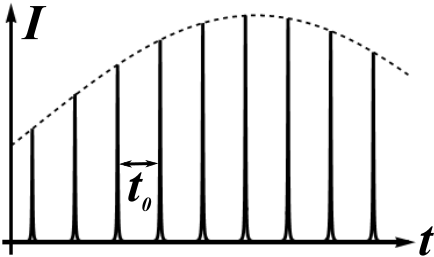}\hspace{1.2cm}

\caption{\label{FigPulses}\small (a) Diagram for the model of linear absorption in two-level system. (b) Intensity of the radiation $I(t)$ is a regular series of $\delta$-pulses separated by time intervals $t_0$. Dashed curve shows the envelope of the pulse intensity.}
\end{figure}

 \textbf{Excitation by intensity modulated light pulses.}~~In case of the intensity modulation of light pulses (see Fig.~\ref{FigPulses} a, b), 
the intensity, $I(t)$, can be represented by the series  
$$
I(t)=I_0\sum_n a_n \delta(t-t_0 n),
$$ 
where $a_n=1+A_I\sin(\phi t)$ (see Eq.\eqref{intensitymodulation}), and the pulses are approximated by a train of delta pulses arriving at time $t_0 n$, where $t_0$ is the time interval between two consecutive pulses and $n$ is an integer. The modulation frequency $\phi$ is chosen such that $\phi < 1/t_0$. 
In what follows, it is convenient to rescale all characteristic times of the problem in by $t_0$. To this end, we introduce a dimensionless time
$$
\sigma\equiv t/t_0,
$$
such that the pulses arrive at each integer value of $\sigma$. We assume that the sample is an ensemble of identical two-level systems placed into a homogeneous surrounding. Let $P(\sigma)$ describe the fraction of systems in the excited state $S_1$. The time evolution of $P(\sigma)$ in our model is given by the kinetic equation ($\Gamma=t_0/\tau$)
\begin{equation}\label{pulsedecay}
\frac{dP(\sigma)}{d \sigma}=-\Gamma P(\sigma)+R(1- P(\sigma))\sum_n a_n \delta(\sigma- n),\qquad P(0)=0.
\end{equation}
 
The properties of $\delta$ function allows us to integrate the above equation and reduce it to the recurrence
\begin{equation}\label{recurence}
P_{n}=\gamma P_{n-1} + R_0a_n(1-\gamma P_{n-1}),\qquad P_0=0,\qquad \gamma=e^{-\Gamma},
\end{equation}
where $\gamma$ is the fraction of the population that remains in the state $S_1$ after the time interval $t_0$ and $R_0 = R \Delta t$ ($\Delta t$ is the pulse duration). The recurrence relation \eqref{recurence} has a simple physical interpretation: The population $P_n$ of the state $S_1$ taken at the instance right after the $n^{\textrm{th}}$ pulse, $P_n=P(n+0)$, is the sum of the population left after the $(n-1)^{\textrm{th}}$ pulse reduced by the factor $\gamma$ and the part of population excited from the ground state $S_0$. The excitation is proportional to the ground state population, $(1-\gamma P_{n-1})$, and the amplitude of the pulse $a_n$ taken with the transition probability $R_0$. The function $P(\sigma)$ has a saw-tooth shape, whose analytic expression is
\begin{equation}\label{pulsedecaySolution}
P(\sigma)=\sum_{n=0}^\infty P_{n}\Omega_n(\sigma)e^{-\Gamma (\sigma-n)},
\end{equation}
where $P_n$ satisfies Eq.~(\ref{recurence}) subject to the initial condition $P_0=0$ and $\Omega_n(\sigma)$ is an indicator of the time-interval $[n ,n+1)$,
$$
\Omega_n(\sigma)=\begin{cases}1,&n \le \sigma<n+1 ;\\
0,&\mbox{otherwise.}
\end{cases}
$$

Note, that Eq.~(\ref{pulsedecay}) in comparison with Eq.~(\ref{continiousdecay}) includes an additional term $I(\sigma)\big(1-P(\sigma)\big)$, which allows us to take into account the non-equilibrium state of the system between the pulses.  
 In our formulation, we have assumed that the total number of photons incident on the sample during the pulse time is small, i.e. $R_0\ll 1$, so that the $S_0$ population is always larger than the $S_1$ population. Therefore, the light-matter interaction is described by a product of the number of systems in $S_0$ and the number of photons. In the opposite case of high intensities, i.e. when $R_0 \sim 1$, each light-pulse can bleach the ground state $S_0$ and the second term in the recurrence~(\ref{recurence}) can get nonphysical negative values. This situation requires a different formulation of the problem that allows the inversion of population in the states $S_0$ and $S_1$.

Let us first consider the case of time-independent radiation intensity with $a_n=1$. The solution of Eq.~(\ref{recurence}) is given by the formula
$$
P_n\Big|_{a_n=1}=\frac{R_0}{1-\gamma+\gamma R_0}\big(1-\gamma^n(1- R_0)^n\big).
$$
As one can see, the population exponentially converges to the steady-state solution $P_{\mathrm{steady-state}}=R_0(1-\gamma+\gamma R_0)^{-1}$. As $a_n$ represents a periodic process, the steady-state follows the same periodicity. Thus, we can ignore the fast transition to the steady-state and start the series in Eq.~(\ref{pulsedecaySolution}) from $n=-\infty$. By taking the Fourier transform of Eq.~(\ref{pulsedecaySolution}) using the formula given in Eq.~(\ref{Fourier}), we get
\begin{multline}\label{Fourierrec}
\mathcal{F}[P(\sigma)](\omega)=\frac{1}{2\pi}\int_{-\infty}^{\infty}P(\sigma)e^{-\i \omega \sigma}d\sigma
=\frac{1}{2\pi}\sum_{n=-\infty}^{\infty} P_{n} \int_{n}^{n +1}e^{-\Gamma (\sigma-n)}e^{-\i \omega \sigma} d\sigma\\=
\frac{1}{2\pi}\frac{\Gamma -\i\omega}{\Gamma^2 +\omega^2}\sum_{n=-\infty}^{\infty}\left(P_n-\gamma P_{n-1} \right) e^{- \i \omega n}.
\end{multline} 

PL from molecular systems is proportional to  the population of the excited state. Hence, for the time-independent radiation (the case when $a_n=1$) the PL contains only the dc component and $\mathcal{F}[P_{\mathrm{steady-state}}](\omega)=R_0(1-\gamma)(1-\gamma+\gamma R_0)^{-1}\delta(\omega/\Gamma)$. In more interesting case, when the pulse intensity is  modulated sinusoidally, i.e. $$a_n=1+A_I\sin( \tilde \phi n),\qquad \tilde \phi =\phi t_0,$$ and the life-time is rather short ($\gamma\ll 1$), one can replace the combination $\left(P_n-\gamma P_{n-1} \right)$ in Eq.~(\ref{Fourierrec}) by $a_n$ (it follows from the above assumptions and the recurrence Eq.~(\ref{recurence})) to reproduce the results Eq.~(\ref{fluorescencemodulation}),~(\ref{amplitudescontinious}) and~(\ref{Fourier}). 

In case of finite, but small $\gamma$ the form of Eq.~(\ref{Fourierrec}) allows us to find a general expansion over $\gamma$. The recurrence Eq.~(\ref{recurence}) together with Eq.~(\ref{Fourierrec}) yields
\begin{multline}\label{Main}
\mathcal{F}[P(\sigma)](\omega)=\frac{R_0}{2\pi}\frac{\Gamma -\i\omega}{\Gamma^2 +\omega^2}\sum_{n=-\infty}^{\infty}\big(a_n-\gamma R_0 a_n a_{n-1}+\gamma^2 R_0 (R_0 a_n a_{n-1} a_{n-2}-a_n a_{n-2})+\dots\big) e^{- \i \omega n}
\end{multline}
It is clear that the presence of the term $I(\sigma)(1-P(\sigma))$ in the model, Eq.~(\ref{pulsedecay}), is responsible for generation of the $a_n a_{n-1}\dots a_{n-k}$ terms, which describe correlations in the modulated amplitudes. This, in turn, modulates the PL intensity at higher integer multiples of~$\phi$.

\begin{figure}[t]
 \begin{center}
\hspace{10pt} \includegraphics[scale=0.12]{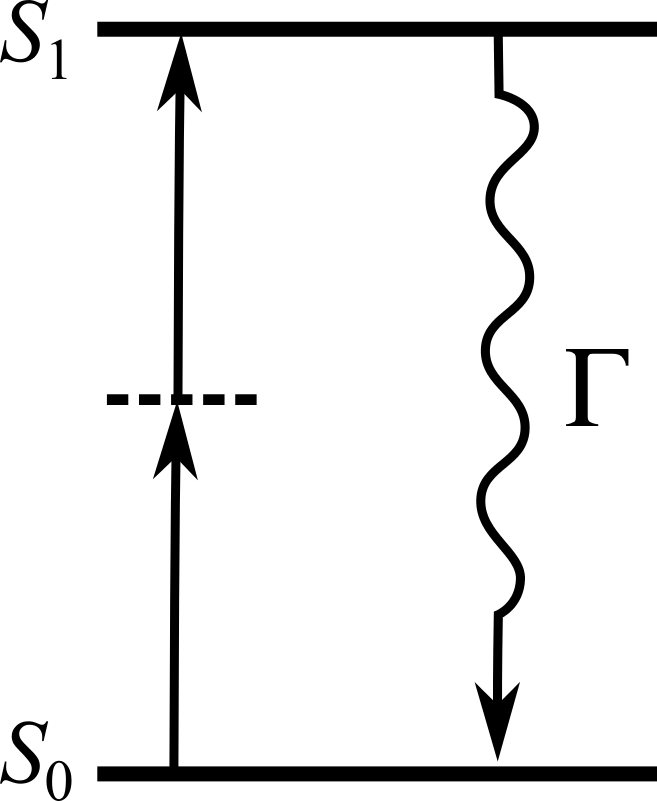}\hspace{50pt} 
 \end{center}
\caption{\label{FigTwoLevelDiag}\small Diagram for the model of two-photon absorption in two-level system.
}
\end{figure}
\textbf{PL from two-photon absorption of modulated light pulses in two-level system.}~~In case of two-photon absorption (Fig.2) the rate is proportional to the square of intensity, so that~\cite{KARKI_2016A}  
\begin{equation}
R = s_2 I_0^2 ,
\end{equation}
where $s_2$ is the two-photon absorption cross-section, and 
\begin{equation}\label{an}
a_n=(1+\cos\tilde{\phi} n)^2 .
\end{equation}
The steady-state solution of the recurrence Eq.~(\ref{recurence}) in the leading order over small $\tilde{\phi}$ is given~by
\begin{equation}\label{ststtwo}
P_n=\frac{\beta a_n}{1 +\beta a_n}, \qquad \textrm{with}\qquad \beta=\frac{\gamma R_0}{1-\gamma}.
\end{equation}
One can calculate also a correction to the steady state solution, which is given by
$$
-\beta\frac{1-R_0a_n}{(1 +\beta a_n)^2}\frac{a_n-a_{n-1}}{1-\gamma}+\mathcal{O}(\tilde{\phi}^2)
$$  
The correction becomes essential only if $1-\gamma\sim\beta\tilde{\phi}$, so that the ratio $\frac{\beta(a_n-a_{n-1})}{1-\gamma}$ is finite. In terms of the life-time $\tau$, we have $\tau\sim 1 /\sqrt{R \tilde{\phi}}$. For low excitation ($\beta\ll 1$) the correction can be neglected and the approximation of the steady state behavior given by the Eq.~(\ref{ststtwo}) works well.

The PL is proportional to the population of $S_1$. The Fourier transform of the signal (in the steady-state approximation) is
\begin{equation}\label{Main1}
\mathcal{F}[P(\sigma)](\omega)=
\frac{R_0(1-\gamma e^{-\i\omega})}{2\pi}\frac{\Gamma -\i\omega}{\Gamma^2 +\omega^2}\sum_{n=-\infty}^{\infty}\frac{a_n}{1 +\beta a_n} e^{- \i \omega n}.
\end{equation}
Since the function $a_n$ is periodic  and depends only on the exponent $e^{\i\tilde{\phi n}}$, the result is a series of $\delta$-functions of the form $\delta(\omega- m \tilde{\phi})$, $m=0,\pm 1,\pm 2,\dots$ For our purposes (note that we only compare the amplitudes at certain peaks) it is enough to consider the positive $m$ only. When $\beta\ll 1$ one can use the Taylor series expansion to approximate the fraction $\frac{a_n}{1 +\beta a_n}$, which gives 
\begin{equation}\label{Main2}
\mathcal{F}[P(\sigma)](\omega\sim m \tilde{\phi})=
\frac{R_0(1-\gamma e^{-\i\omega})}{2\pi}\frac{\Gamma -\i\omega}{\Gamma^2 +\omega^2}\sum_{n=-\infty}^{\infty}(a_n -\beta a_n^2+\beta^2 a_n^3+\dots) e^{- \i \omega n}.
\end{equation}
The modulation frequencies of the PL intensity calculated by Eq.~(\ref{Main2}) up to the second order in $\beta$ yields (only essential terms are presented)
\begin{multline}
\mathcal{F}[P(\sigma)](\omega)=
R_0\frac{\Gamma -\i\tilde{\phi}}{\Gamma^2 +\tilde{\phi}^2}(1-\gamma e^{-\i\tilde{\phi}})\left(1-\frac{7}{2}\beta+\frac{99 }{8}\beta^2\right)\delta(\omega-\tilde{\phi}) \\ +\frac{R_0}{4}\frac{\Gamma -\i2\tilde{\phi}}{\Gamma^2 +4\tilde{\phi}^2}(1-\gamma e^{-\i 2\tilde{\phi}})\left(1-7\beta+\frac{495}{16}\beta^2\right)\delta(\omega-2\tilde{\phi})\\ -\frac{R_0\beta}{2}\frac{\Gamma -\i3\tilde{\phi}}{\Gamma^2 +9\tilde{\phi}^2}(1-\gamma e^{-\i 3\tilde{\phi}})\left(1-\frac{55}{8}\beta\right)\delta(\omega-3\tilde{\phi})\\-\frac{R_0\beta}{16}\frac{\Gamma -\i4\tilde{\phi}}{\Gamma^2 +16\tilde{\phi}^2}(1-\gamma e^{-\i 4\tilde{\phi}})\left(1-\frac{33}{2}\beta\right)\delta(\omega-4\tilde{\phi})+\mathcal{O}(\beta^3)+\dots\nonumber
\end{multline}
Thus the ratio of the amplitudes $A_{m\phi}$ that appear at the peaks up to the second order approximation in $\beta$ are 
\begin{eqnarray}
\frac{A_{2\phi}}{A_{\phi}}&=&\frac{1}{4}\sqrt{\frac{\Gamma^2+4\tilde{\phi}^2}{\Gamma^2+\tilde{\phi}^2}\cdot\frac{1-2\gamma\cos 2\tilde{\phi} +\gamma^2}{1-2\gamma\cos \tilde{\phi} +\gamma^2}}\left(1-\frac{7}{2}\beta+\frac{101}{16}\beta^2\right),\label{16}\\
\frac{A_{3\phi}}{A_{\phi}}&=&\frac{\beta}{2}\sqrt{\frac{\Gamma^2+9\tilde{\phi}^2}{\Gamma^2+\tilde{\phi}^2}\cdot\frac{1-2\gamma\cos 3\tilde{\phi} +\gamma^2}{1-2\gamma\cos \tilde{\phi} +\gamma^2}}\left(1-\frac{27}{8}\beta\right),\label{17}\\
\frac{A_{4\phi}}{A_{\phi}}&=&\frac{\beta}{16}\sqrt{\frac{\Gamma^2+16\tilde{\phi}^2}{\Gamma^2+\tilde{\phi}^2}\cdot\frac{1-2\gamma\cos 4\tilde{\phi} +\gamma^2}{1-2\gamma\cos \tilde{\phi} +\gamma^2}}\left(1-13\beta\right).\label{18}
\end{eqnarray}
In the limiting case when  the life-time of $S_1$ state is much shorter than $t_0$,   $\Gamma \gg \tilde{\phi}$ and $\beta\approx 0$, which gives $A_{2\phi}:A_{\phi}\approx 1:4$. In other cases, when $\beta$ is non-negligible, the ratio $A_{2\phi}:A_{\phi}$ is smaller than 1:4. Note, the finite life-time of the photoluminescence ($\beta\ne 0$) gives non-zero amplitudes at frequencies at higher multiples of $\phi$. 

\begin{figure}
 \begin{center}
  \includegraphics[scale=0.14]{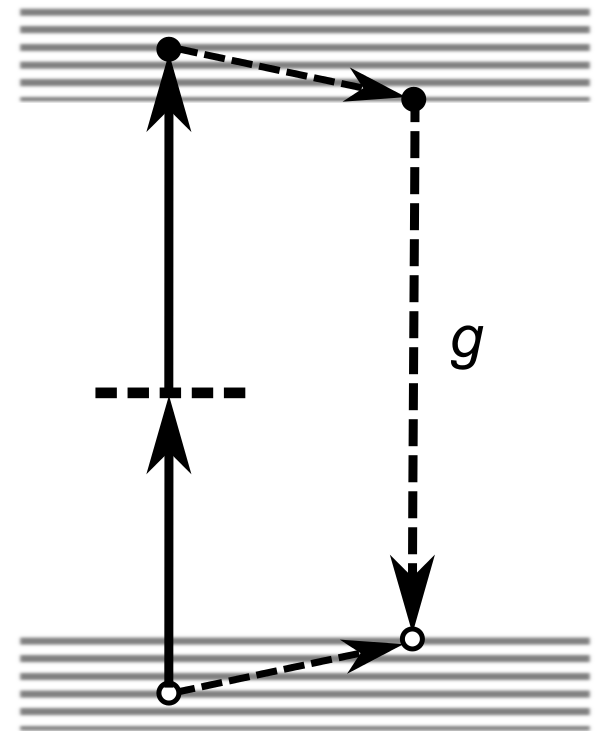}
 \end{center}
\caption{\label{FigSemi}\small Model of two-photon absorption of the modulated light pulses in semiconductor. }
\end{figure}

\textbf{Photocurrent from two-photon absorption in semiconductors.}\label{SecSemi}~~
The preceding derivations can also be extended to non-exponential relaxation processes, such as the recombination processes in semiconductors. Although there are a number of processes by which the electrons and the holes in a semiconductor can recombine,~\cite{L1991} we consider only the case of the band-to-band recombination of the charge carriers (Fig.\ref{FigSemi}). 
Let $P(t)$ denote the concentration of electrons (holes) in the semiconductor. When the rates of creation and annihilation are in a balance at equilibrium, the product of the electron and hole densities is a constant ($P^2(t)=const$). The thermal excitation of electrons in a large band-gap semiconductor is very small, so the corresponding constant is irrelevant in further analysis. The charge carriers generated by the two-photon absorption results in the increase of electron-hole pairs, the reverse process of band-to-band recombination is quadratic in $P$, thus
\begin{equation}\label{semieq}
\frac{dP(\sigma)}{d\sigma}=-gP^2(\sigma)+R\sum_n a_n \delta(\sigma- n),\qquad P(0)=0,
\end{equation}
where $g$ is the recombination rate.
The solution of Eq.~(\ref{semieq}) is
\begin{eqnarray}\label{PDECAY}
P(\sigma)&=& \sum_{n=0}^\infty \Omega_n(\sigma)\frac{P_n}{1+g (\sigma-n) P_n},
\end{eqnarray}
where $P_n$'s satisfy the recurrence
\begin{eqnarray}\label{recurrencesemi}
P_{n+1}=R_0a_n+\frac{P_n}{1+g P_n}
\end{eqnarray}

The experimental observable is the photocurrent, which is proportional to the number of free carriers generated in the media. The Fourier transform of the observed signal is then given by the formula
\begin{equation}\label{SemiFourier}
\mathcal{F}[P(\sigma)](\omega)=\frac{1}{2\pi}\sum_{n=-\infty}^{\infty} e^{-\i \omega n} \int_0^1\frac{P_n}{1+g \sigma P_n}e^{-\i \omega \sigma} d\sigma
\end{equation}
As soon as the Fourier transform is calculated in the vicinity of $m\tilde{\phi}$ one can omit the  exponent under the integral, so that $\mathcal{F}[P(\sigma)](\omega\sim m\tilde{\phi})=\frac{1}{2\pi g}\sum_{n=-\infty}^{\infty}\log(1+g P_n) e^{-\i \omega n}$.
The steady state solution of the recurrence~(\ref{recurrencesemi}) has the form $
P_n=\frac{R_0}{2}\left( a_n+\sqrt{a_n^2+\frac{4 a_n}{gR_0}}\right)
$. Its substitution into the above expression leads to a simple expression
\begin{equation}\label{FourierSemi}
\mathcal{F}[P(\sigma)](\omega\sim m\tilde{\phi})=\frac{1}{2\pi g}\sum_{n=-\infty}^{\infty}\log\left(1+\mu\left( a_n+\sqrt{a_n^2+\frac{2 a_n}{\mu}}\right)\right) e^{-\i \omega n},\qquad \mu=\frac{gR_0}{2}.
\end{equation}

As one can see, $\mu$ is the main governing parameter of the model. We analyze the behavior of $A_{2\phi}:A_{\phi}$ at two limiting cases, namely when $\mu \ll 1$ and $\mu \gg 1$. In the case when $\mu \ll 1$, $\mu\left( a_n+\sqrt{a_n^2+\frac{2 a_n}{\mu}}\right) \approx \mu\left( a_n+\sqrt{a_n} \sqrt{\frac{2}{\mu}}\right)\leq 1 $, and the logarithm can be expanded using Taylor's series to obtain the ratio $A_{2\phi}: A_{\phi}$, which is given by 
\begin{equation}\label{CLR}
A_{2\phi}:A_{\phi} = \frac{\mu^{\frac{3}{2}}(112-1200\sqrt{2\mu}-...)}{64\sqrt{2}-80\sqrt{2} \mu+...}.
\end{equation}
As $\mu \rightarrow 0$, the ratio also goes to zero. The ratio demonstrates a few oscillations at small values of $\mu$, while in the limit $\mu \gg 1$, when $\mu\left( a_n+\sqrt{a_n^2+\frac{2 a_n}{\mu}}\right) \approx 2\mu a_n$, the logarithm can be approximated by 
\begin{equation}\label{CRB1}
\log\left(1+\mu\left( a_n+\sqrt{a_n^2+\frac{2 a_n}{\mu}}\right)\right)\approx \log (2)+\log (\mu)+\log (a_n). 
\end{equation}  
In Eq.\eqref{CRB1}, only the term $\log(a_n)$ contains modulated signal, which can be written as 
  \begin{equation}\label{CRB2}
  \log(a_n)= \log(2)-\frac{1}{4}+\cos(\tilde{\phi}n)-\frac{1}{4}\cos(2\tilde{\phi}n)+...
    \end{equation}    
    Thus, the ratio of the amplitudes at large values of $\mu$ approaches the value $1:4$. Analysis shows, that this part of the curve is universal (independent on choice of $\phi$).

\subsection{Experimental setup}\label{SecExp}
\begin{figure}
  \begin{center}
\includegraphics[width=0.9\textwidth]{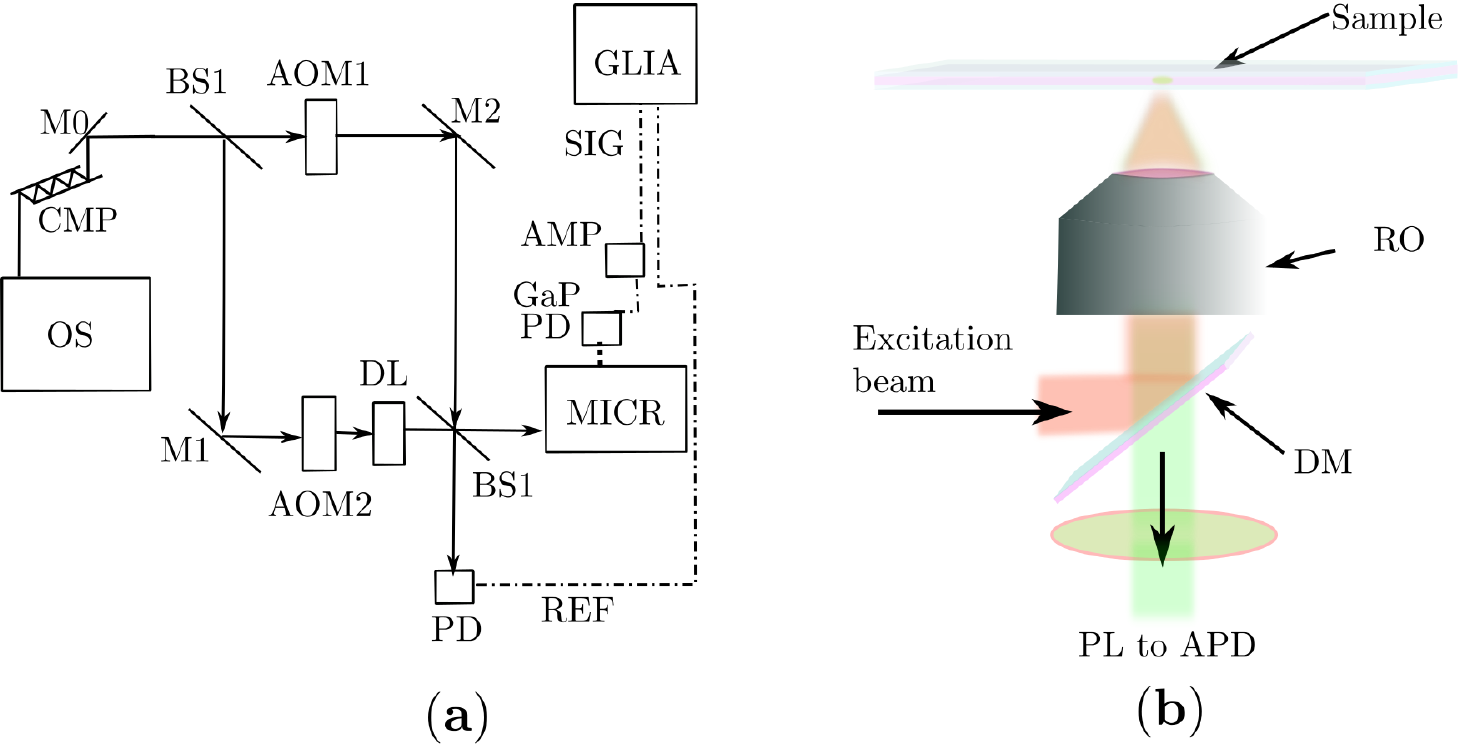}
 \end{center}
\caption{\label{FigExp}\small  (a) Schematics for the experimental setup used to measure two-photon photocurrent from a GaP photodiode. OS: mode-locked oscillator, CMP: chirp-mirror pair, M: mirror, BS: beam splitter, AOM: acousto-optic modulator, DL: delay line, PD: photodiode, REF: reference, MICR: microscope (inverted), AMP: current-to-voltage amplifier, SIG: signal and GLIA: generalized lock-in amplifier. (b) Schematic of focusing optics in microscope used in excitation and collection of two-photon photoluminescence from rhodamine 6G. DM: dichroic mirror, PL: photoluminescence, APD: avalanche photodiode, RO: reflective objective (cassagrain).}
\end{figure}

The schematic of the optical setup used in the measurements of the two-photon photocurrent from GaP photodiode is shown in Fig.~\ref{FigExp}(a). Briefly,  a Ti:Sapphire oscillator (Synergy from Femtolasers, center wavelength 780 nm, bandwidth 135 nm, repetition rate 73MHz) was used as the optical source (OS). A pair of chirp mirrors (CMP, Layertec, part no. 111298) were used to compensate the group velocity dispersion induced by the different dispersive optical elements. A 50/50 beam splitter (BS1) was used to split the beam from OS into two identical replicas. The phases of each of the beams were modulated by acousto-optic modulators (AOMs) placed on the arms of a Mach-Zehnder interferometer. The phase modulation frequencies were set to $\phi_1=54.7$ MHz and $\phi_2 = 54.75$ MHz, respectively, for AOMs 1 and 2, such that the difference in the phase modulation was $\phi_{21}=0.05$ MHz. A piezo driven retroreflector (DL) in one of the arms of the interferometer was used to adjust the optimal temporal overlap between the two beams. A second beam splitter (BS2) combined the two beams. One of the outputs from BS2 was sent to the microscope, while the other output, which was monitored by a photodiode (PD), served as the reference. We used an inverted microscope (Nikon Ti-S) in the setup. A reflective objective (RO, 36X/0.5 NZ, Edmund Optics) was used to focus the beam onto a gallium phosphide photodiode (Thorlabs, part. no. FGAP71). The size of the focus spot was about 1.5 $\mu$m. The pulse duration, $\Delta t$, at the sample was about 10 fs. The photocurrent from the photodiode was amplified by an amplifier (SR570, Stanford Research Systems) and the output of the amplifier was analyzed by using a generalized lock-in amplifier (GLIA).~\cite{KARKI_2013A,KARKI_2013C,KARKI_2014B} 

The schematic setup used to measure the two-photon PL from rhodamine 6G in methanol is shown in Fig.~\ref{FigExp}(b). The two-photon PL from a 10 millimolar solution was detected in the epi-direction using the same microscope objective. A dichroic mirror (DM, FF670-SDi01-25x36) was used to separate the PL from the excitation beam. Another band pass filter (FIL, 550 nm X 50 nm, OD 5, Edmund Optics, part no. 84772) was used to further suppress the scattered light from reaching the detector. An avalanche photodiode with a bandwidth of 2 MHz (APD, Laser Components, part no. LCSA3000-01) was used to detect the PL. As in case of the photocurrent detection scheme, the final electronic signal from the APD was digitized by a digitizer and analyzed using a generalized lock-in amplifier.

\section{Results and discussion}

\begin{figure}
 \begin{center}
 \includegraphics[scale=0.7]{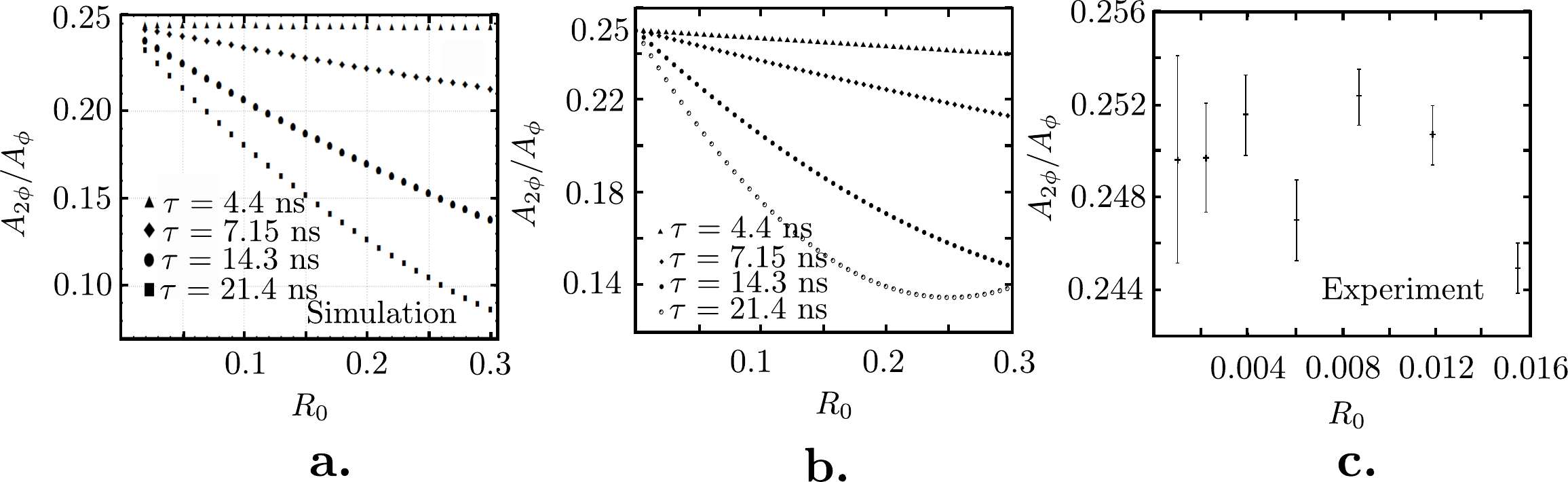}
 \end{center}
\caption{\label{FigTwoLevelApprox}\small  (a) The ratio of amplitudes $A_{2\phi}:A_{\phi}$ calculated by the formulae~(\ref{recurence}),~(\ref{Fourierrec}) are plotted with respect to $R_0$ for various $\tau$, all $\tau$ are measured in $t_0$ and $\phi=1/6$; (b) the amplitude ratios $A_{2\phi}:A_{\phi}$ calculated by using the approximate formula~(\ref{16}); (c) the ratio obtained from the  measurements of two-photon photoluminescence from rhodamine 6G in methanol.}
\end{figure}

\textbf{Two-photon PL from rhodamine 6G.}~~Fig.\ref{FigTwoLevelApprox}(a) shows the ratio of the amplitudes computed by simulating the population of two-level system by using Eqs.\eqref{recurence}--\eqref{pulsedecaySolution} and the Fourier transform of the excited state population using Eq.\eqref{Fourierrec}. As can be seen in the figure, the ratio is 1:4 irrespective of $\tau$ when the transition probability $R_0$ is close to zero. For large $R_0$ and for $\tau$ comparable to $t_0$, the ratio gradually decreases from 1:4. This behavior is well reproduced by the approximate formulae Eq.\eqref{16}--\eqref{18} at lower values of $R_0$ (Fig.\ref{FigTwoLevelApprox} (b)), and the deviations at higher values of $R_0$ can be minimized by taking into account the higher order corrections.

On Fig.\ref{FigTwoLevelApprox} (c), we show the experimentally determined ratio of $A_{2\phi}:A_{\phi}$ in rhodamine 6G at different transition probability $R_0$. $R_0$ is varied by changing the intensity of the laser beams. Under two-photon excitation, the transition probability is given by Eq.\eqref{R_0_REL}
\begin{equation}\label{R_0_REL}
R_0 = s_2 I_0^2 \Delta t.
\end{equation} 
The two-photon absorption cross-section, $s_2$, of rhodamine 6G is about $15\times 10^{-50}$ cm$^4$ s photon$^{-1}$.~\cite{GOSWAMI_2009}
At the lowest and the highest excitation intensities, we have pulse energies of 0.034 nJ and 0.136 nJ, respectively.  The corresponding transition probabilities are $R_0=9.7\times 10^{-4}$ and $1.55\times 10^{-2}$, respectively. In our experiments, the lowest excitation intensity is limited by the detector sensitivity and the highest one by the maximum available laser power. Note that in both cases we are in the regime where $R_0\ll 1$.  The PL life-time of rhodamine 6G in methanol at 10 millimolar concentration is about 4.4 ns,~\cite{SIKKELAND_1977} which is almost three times smaller than the time interval between the laser pulses ($t_0 \simeq 14.29$ ns). Under these conditions, $\Gamma\approx 3.24$, $\gamma \approx 0.04$, $\tilde{\phi}=7.145\times10^{-4}$ and $4\times10^{-5}\leq\beta\leq 6.46\times 10^{-4}$. Using these parameters in Eq.\eqref{16}, we get $A_{2\phi}:A_{\phi} \approx 1:4$ for all the excitation intensities in our measurements, which agrees with the results from the full simulation using Eq.\eqref{recurence} and  Eq.\eqref{Fourierrec} and the approximate formula given by Eq.\eqref{16}.

\begin{figure}
 \begin{center}
 \includegraphics[scale=0.7]{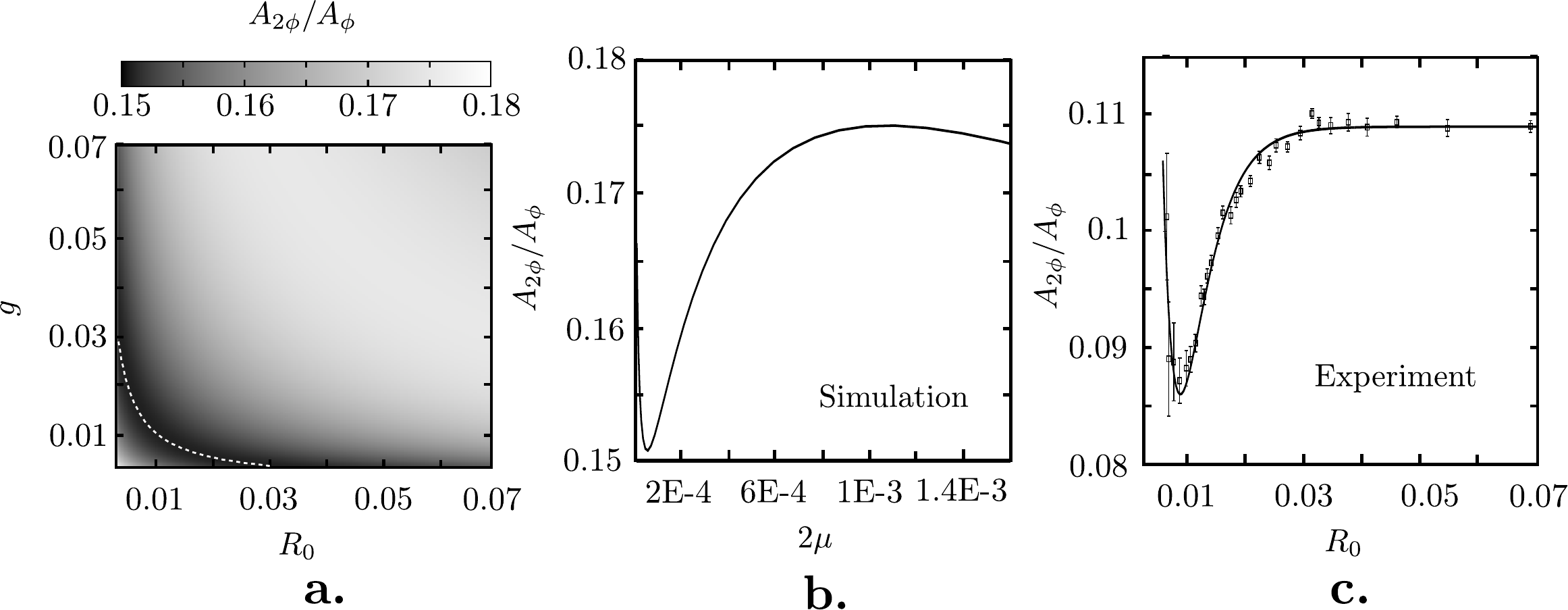} 
\end{center}
\caption{\label{FigGAP}\small  (a) The ratio of amplitudes $A_{2\phi}:A_{\phi}$   as a function of $g$ and $R_0$ obtained from the simulations, (b) the ratio as a function of $gR_0=2\mu$ and (c) the experimentally determined ratio in GaP as a function of $R_0$ (solid line here is a spline interpolation).}
\end{figure}

\textbf{Two-photon photocurrent from GaP.}~~ We have used the recombination kinetics (Eq.\eqref{PDECAY}) and the recurrence relation (Eq.\eqref{recurrencesemi}) to simulate the concentration of electrons in the conduction band of GaP as a function of time. The photocurrent is proportional to the concentration of free electrons (or charge carriers). The ratio $A_{2\phi}:A_{\phi}$ in the photocurrent as a function of $R_0$ and $g$ is shown in Fig.\ref{FigGAP}(a). The two-photon absorption cross-section of GaP is about $s_2\approx 1.2\times 10^{-48}$ cm$^4$ s photon$^{-1}$.~\cite{MINAFRA_1975} The photon flux at the lowest and the highest excitation densities we have used in our experiments are $2.2\times 10^{29}$ photons cm$^{-2}$ s$^{-1}$ and $2.4\times 10^{30}$ photons cm$^{-2}$ s$^{-1}$, respectively. The corresponding transition probabilites for the highest and the lowest excitation densities are 5.8$\times 10^{-4}$ and $6.9\times 10^{-2}$, respectively. Thus, in Fig.\ref{FigGAP}, we have only shown the ratio for $R_0$ that correspond to the experimental conditions.

The ratio calculated from the simulations show that it is a function of the product $gR_0$, which is in agreement with Eq.\eqref{FourierSemi}. Fig.\ref{FigGAP}(b) shows the ratio along the diagonal as a function of $2\mu=gR_0$. In Fig.\ref{FigGAP}(c), we show the experimentally observed ratio as a function of $R_0$. The measurements are done by varying the excitation intensity and the corresponding $R_0$ are calculated by using the Eq.\eqref{R_0_REL}. As it can be seen, the ratio has a minimum at $R_0=0.009$, similar to the one in Fig.\ref{FigGAP}(b) at $2\mu = 1.17\times 10^{-4}$. Comparing Fig.\ref{FigTwoLevelApprox} and Fig.\ref{FigGAP}, we observe that the first order and the second order relaxation processes can be distinguished qualitatively simply by the dependence of the ratio on $R_0$. In a first order process the ratio decreases monotonically with increasing $R_0$, while in a second order process we observe undulations.  Moreover, in the second order relaxation process, we can accurately estimate the value of the recombination rate $g$ from the position of the minimum. Analyzing the data presented on the Fig.\ref{FigGAP} (b) and (c), we get $g=0.013$, which is the number of recombinations during the interval between the pulses. In the units of recombinations per second, the rate is $g/t_0 \approx 9.1\times 10^5$ s$^{-1}$. It is customary to report the recombination parameters in terms of recombination coefficient, which we calculate as follows. The recombination coefficient is given by $r = g/\rho_c$, where $\rho_c$, is the density of free charge carriers created by the absorption of photons. The energy absorbed during the two-photon excitation process is given by $\Delta E = \Delta I  \Delta t\ominus $, where $\Delta I$ is the change in the intensity of the light after absorption and $\ominus$ is the area of the focus spot. The change in the intensity can be calculated by the relation $\Delta I = s_2 x I_m^2/(V E)$, where $V$ is the volume of the unit cell of GaP, $E$ is energy of a single photon, $x$ is twice the Rayleigh length of the beam (confocal parameter), and $I_m$ is the intensity corresponding the $R_0=0.009$, i.e, $R_0$ at the minima. Using the following values: $s_2 = 1.2\times 10^{-48}$ cm$^4$ s photon$^{-1}$, $V=1.62\times 10^{-21}$ cm$^3$, $x=4.4\times 10^{-4}$ cm, $I_m=2.23\times 10^{11}$ W cm$^{-2}$ and $E=2.48\times 10^{-19}$ J photon$^{-1}$, we get $\Delta I = 6.56\times 10^{10}$ W cm$^{-2}$ and $\Delta E = 11.6  \times 10^{-12}$ J. The number of absorbed photons is $\Delta n = 4.7\times 10^7$, and the number of generated electron-hole pairs is $n_c = \Delta n/2 = 2.35\times 10^7$. Then, the density of the free charge carriers is $\rho_c \approx 0.3\times 10^{19}$ cm$^{-3}$, which gives the recombination coefficient $r\approx 3\times 10^{-13}$ cm$^3$ s$^{-1}$. The recombination coefficient in GaP measured by using other optical techniques is about $10^{-13}$ cm$^3$ s$^{-1}$,\cite{DEAN_76} which is similar to the value obtained above.    

\section{Conclusions and outlook}
We have presented theoretical analysis of the actions signals, PL from molecules and photocurrent from semiconductors, that were detected from intensity modulated pulsed lasers. Although, it has been widely accepted that the excitation at a fixed modulation frequency $\phi$  modulates the action signals at the same frequency\cite{LAKOWICZ_2006}, our analysis shows that the action signals can have modulations at higher frequencies, $n\phi$, if the lifetime of the signals is longer than the repetition rate of the laser. Our analysis of the two-photon PL shows that ratio of the signals modulated at $2\phi$ and $\phi$ depends on the excitation intensity. At low excitation intensity, the ratio converges to the previously known values,\cite{XU_2013} and at higher excitation intensity the ratio decreases (until it goes up again for extremely high intensities, the regime hardly achievable in experiment). Moreover, when the lifetime of the pholuminescence is short compared to the repetition rate of the laser the ratio becomes 1:4. We also show that the intensity dependence of the ratio of the signals at $2\phi$ and $\phi$ can used to distinguish different relaxation processes. Although, the ratio decreases monotonically with increasing intensity for the first order relaxation process, it shows undulations in the case of a second order relaxation process. We have experimentally verified these theoretical predictions. More importantly, we have also shown that the recombination coefficient of free charge carriers in semiconductors can be quantified using the ratio. Our calculations show that the recombination coefficient in a GaP photodiode is about $3\times 10^{-13}$ cm$^3$s$^{-1}$.

In this article, we have  analyzed only the action signals due to the first and the second order relaxation processes. Contributions of other higher order relaxation processes, such as Auger recombination, on the ratio of the signals at $2\phi$ and $\phi$ as a function of the excitation intensity remain to be investigated. Moreover, the ratio could also be sensitive to spatial dynamics of excitations, such as energy transfer, diffusion and drift of free charges. The analysis we have followed in this article can be generalized to include such processes too.

\section*{Acknowledgements} This work was supported by the research grants from Swedish Research Council (VR), Crafoord Foundation and NonoLund.

\end{document}